\documentclass[journal, 11pt, onecolumn]{IEEEtran}
\IEEEoverridecommandlockouts
\usepackage{graphicx}
\usepackage{caption}
\usepackage{subcaption}
\usepackage{mathtools}
\usepackage{bm,amsmath, amssymb}
\usepackage{amsmath}
\usepackage{amsfonts}
\usepackage {amssymb,amsmath}
\usepackage{cite}
\usepackage{epstopdf}
\usepackage{color}
\usepackage{microtype}
\DeclareGraphicsExtensions{.pdf,.png,.jpg}

\usepackage{cite}

\def\BibTeX{{\rm B\kern-.05em{\sc i\kern-.025em b}\kern-.08em
    T\kern-.1667em\lower.7ex\hbox{E}\kern-.125emX}}
    
\begin{document}
\title{{Waveform for Next Generation Communication Systems: Comparing Zak-OTFS with OFDM}}
\author{\IEEEauthorblockN{Imran Ali Khan\IEEEauthorrefmark{1}, Saif Khan Mohammed\IEEEauthorrefmark{1},
Ronny Hadani\IEEEauthorrefmark{2}\IEEEauthorrefmark{3},
Ananthanarayanan Chockalingam\IEEEauthorrefmark{4}, \\
Robert Calderbank\IEEEauthorrefmark{5},~\IEEEmembership{Fellow,~IEEE},  Anton Monk\IEEEauthorrefmark{3}, Shachar Kons\IEEEauthorrefmark{3}, Shlomo Rakib\IEEEauthorrefmark{3},
and Yoav Hebron\IEEEauthorrefmark{3}
\thanks{This work has been submitted to the IEEE for possible publication. Copyright may be transferred without notice, after which this version may no longer be accessible.}}

\vspace{5mm}

\IEEEauthorblockA{\IEEEauthorrefmark{1}Department of Electrical Engineering, Indian Institute of Technology Delhi, India}\\
\IEEEauthorblockA{\IEEEauthorrefmark{2}Department of Mathematics, University of Texas at Austin, USA}\\
\IEEEauthorblockA{\IEEEauthorrefmark{3} Cohere Technologies Inc., Santa Clara, CA, USA}\\
\IEEEauthorblockA{\IEEEauthorrefmark{4} Department of Electrical Communication Engineering, Indian Institute of Science Bangalore, India}
\IEEEauthorblockA{\IEEEauthorrefmark{5}Department of Electrical and Computer Engineering, Duke University, USA}}

\maketitle

\begin{abstract}
Across the world, there is growing interest in new waveforms, Zak-OTFS in particular, and over-the-air implementations are starting to appear. The choice between OFDM and Zak-OTFS is not so much a choice between waveforms as it is an architectural choice between preventing inter-carrier interference (ICI) and embracing ICI. In OFDM, once the Input-Output (I/O) relation is known, equalization is relatively simple, at least when there is no ICI. However, in the presence of ICI the I/O relation is non-predictable and its acquisition is non-trivial. In contrast, equalization is more involved in Zak-OTFS due to inter-symbol-interference (ISI), however the I/O relation is predictable and its acquisition is simple. {Zak-OTFS exhibits superior performance in doubly-spread 6G use cases with high delay/Doppler channel spreads (i.e., high mobility and/or large cells), but architectural choice is governed by the typical use case, today and in the future. What is typical depends to some degree on geography, since large delay spread is a characteristic of large cells which are the rule rather than the exception in many important wireless markets.} This paper provides a comprehensive performance comparison of cyclic prefix OFDM (CP-OFDM) and Zak-OTFS across the full range of 6G propagation environments. The performance results provide insights into the fundamental architectural choice.
\end{abstract}

\section{Carrier waveform: Zak-OTFS and CP-OFDM}
A carrier waveform in CP-OFDM is a sinusoid with frequency an integer multiple of the sub-carrier spacing $\Delta f$. The information grid is defined in the frequency domain (FD), the number of  carriers is $B/\Delta f$ where $B$ is the bandwidth, and the time duration $T =  1/\Delta f$. The carriers are pairwise orthogonal, enabling low-complexity per sub-carrier equalization at the receiver.

A Zak-OTFS carrier waveform is a pulse in the delay-Doppler (DD) domain, that is a quasi-periodic localized function defined by a delay period $\tau_p$ and a Doppler period $\nu_p = 1/ \tau_p$. When viewed in the time-domain (TD) this function is realized as a pulse train modulated by a tone, hence the name pulsone. The time duration ($T$) and bandwidth ($B$) of a pulsone are inversely proportional to the characteristic width of the DD domain pulse along the Doppler axis and the delay axis respectively. The number of non-overlapping DD domain pulses, each spread over an area $1/BT$, is equal to the
time-bandwidth product $BT$. In Zak-OTFS modulation, the information carrying DD domain pulses are located at $(\tau, \nu) = \left(k/B, l/T\right)$, $k=0,1,\cdots,M-1$, $l=0,1,\cdots, N-1$ where $M = B \tau_p$ and $N = \nu_p T$, so that there are $MN= BT$ quasi-periodic DD domain pulses, each corresponding to a Zak-OTFS carrier waveform (see \cite{zakotfs1, zakotfs2, otfsbook} for more details).


   \begin{figure}
     \centering
        \vspace{-5mm}
        \includegraphics[width=9.32cm,height=6.0cm]{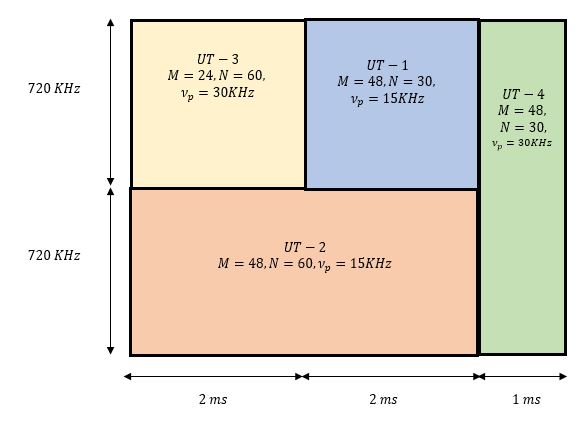}
        \vspace{-2mm}
        \caption{Orthogonal allocation of TF resources.}
        \label{fig1}
        \vspace{-3mm}
    \end{figure}
    
\section{Numerology and overheads}
In both CP-OFDM and Zak-OTFS, it is possible to allocate non-overlapping time-frequency (TF) resources, as is shown in Fig.~\ref{fig1}.

\subsection{The need for cyclic-prefix}
In CP-OFDM, the bandwidth allocated to each user is an integer multiple of the sub-carrier spacing $\Delta f$. The time duration allocated to each user is an integer multiple of the symbol period $T_{prb} = T_{cp} + 1/\Delta f$, where $T_{cp}$ is the cyclic prefix (CP). \emph{The CP needs to be greater than or equal to the channel delay spread} in order to prevent consecutive OFDM symbols from interfering in time. 
   \begin{figure}
     \centering
        \vspace{-1mm}
        \includegraphics[width=9.32cm,height=6.0cm]{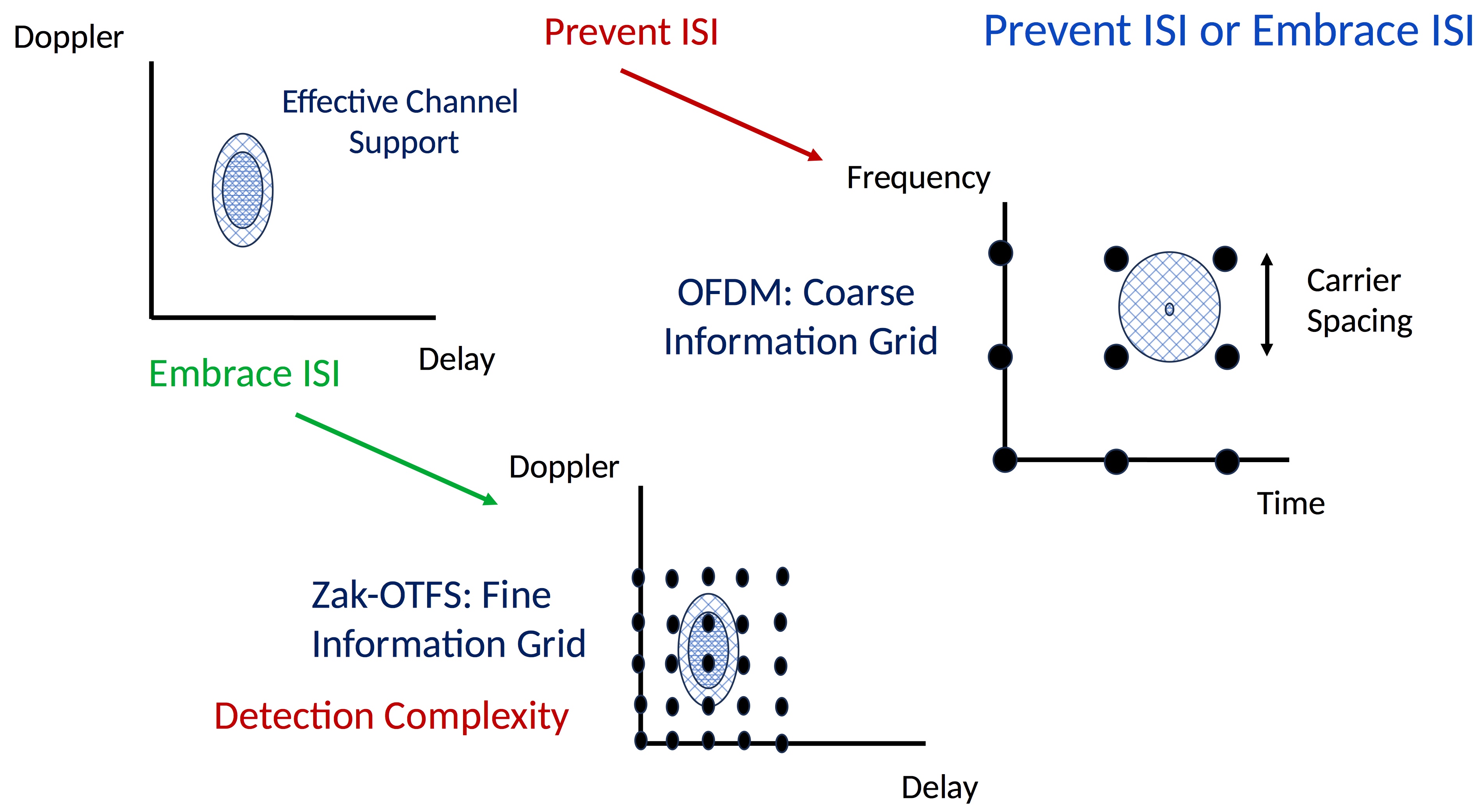}
        \vspace{-2mm}
        \caption{Zak-OTFS (Embracing ICI) vs. CP-OFDM (avoiding ICI).}
        \label{fig2}
        \vspace{-3mm}
    \end{figure}
    
In Zak-OTFS, the time duration and bandwidth allocated to each user are integer multiples of the delay period $\tau_p$ and the Doppler period $\nu_p$ respectively. \emph{The energy of a DD domain pulse is spread throughout the allocated TF resource}; hence the out-of-band leakage is significantly smaller than that of OFDM, and \emph{it is not necessary to provision gaps} in either the TD or the FD. 

\subsection{Acquiring the input-output (I/O) relation}
 {I/O relation refers to the relation between the information-bearing carrier (sub-carrier in the case of OFDM and pulsone in the case of Zak-OTFS) input to the channel and its channel response (output). This I/O relation (in frequency domain for OFDM and DD domain for Zak-OTFS) needs to be acquired for equalization. When there is no ICI (or insignificant ICI due to small time variations in the channel), the I/O relation is predictable. That is, the channel response to a particular OFDM sub-carrier (of some frequency $f_1$) is predictable from the knowledge of response to a given sub-carrier of some other frequency $f_0$. However, when the channel is doubly-spread with significant ICI, the OFDM I/O relation becomes unpredictable, i.e., the channel response to a particular OFDM sub-carrier is difficult to predict from the knowledge of response to a sub-carrier of some other frequency.
Since the channel response to a sub-carrier is the $N_{sub}$-dimensional vector of response received on all $N_{sub} = B/\Delta f$ sub-carriers, it is not possible to acquire the channel response to all sub-carriers even if we transmit pilots on all sub-carriers at the same time (since this is equivalent to learning/estimating $N_{sub}$ $N_{sub}$-dimensional vectors from only a single $N_{sub}$-dimensional vector of response received on all sub-carriers). 
Stated a little differently, one could think of transmitting pilot on only one sub-carrier every OFDM symbol 
and trying to learn the complete response to all sub-carriers over a duration of $N_{sub}$ symbols, 
which is not possible in the presence of channel Doppler spread since the channel response also varies in time.
{ 
Due to these reasons, {\emph{it is not possible to completely acquire the OFDM I/O relation} for a doubly-spread channel.}}}

On the other hand, the I/O relation of a Zak-OTFS pulsone is predictable as long as the delay and Doppler period are respectively greater than the channel delay and Doppler spread (called as the crystallization condition (see {\cite{zakotfs2}}). That is, when the crystallization condition is satisfied,
the channel response to any pulsone can be predicted from the response to a particular pulsone 
(the channel response to a pulsone carrier is the $MN$-dimensional vector of the response received on each of the $MN$ pulsone carriers). This enables a stationary \emph{Zak-OTFS I/O relation in the DD domain which can be completely acquired using a single pulsone}. {This predictability attribute of Zak-OTFS allows a simple and efficient means to acquire the I/O relation in doubly-spread channels.}

\subsection{To prevent ICI or to embrace ICI}
In CP-OFDM the sub-carrier spacing $\Delta f$ is typically chosen to be $10$ times greater than the channel Doppler spread. Also, the symbol duration $T$ is also designed to be significantly larger than the CP duration (i.e., maximum channel delay spread) so that the CP overhead is small. This choice of a \emph{coarse} information grid prevents ICI, it preserves orthogonality of sub-carriers, and it enables low-complexity per sub-carrier equalization at the receiver \cite{Wang2006} (see Fig.~\ref{fig2}).  

In Zak-OTFS the carrier spacing  ($1/B$) in delay is less than the channel delay spread and the carrier spacing in Doppler ($1/T$) is less than the channel Doppler spread. This choice of a fine information grid embraces ICI. The carrier waveforms are designed to interfere with each other in the same way, and joint equalization of all carriers achieves robustness to channel delay / Doppler spread. The price of embracing ICI is higher equalization complexity. Note that \emph{joint equalization is a possibility in Zak-OTFS} as its I/O relation is \emph{predictable and can be acquired completely} whereas joint equalization of all OFDM sub-carriers is \emph{not possible} as its I/O relation cannot be acquired completely.

    \begin{figure}[h!]
    	\vspace{-3mm}
    	\begin{subfigure}[b]{0.5\textwidth}
    	\includegraphics[width=1.0\textwidth]{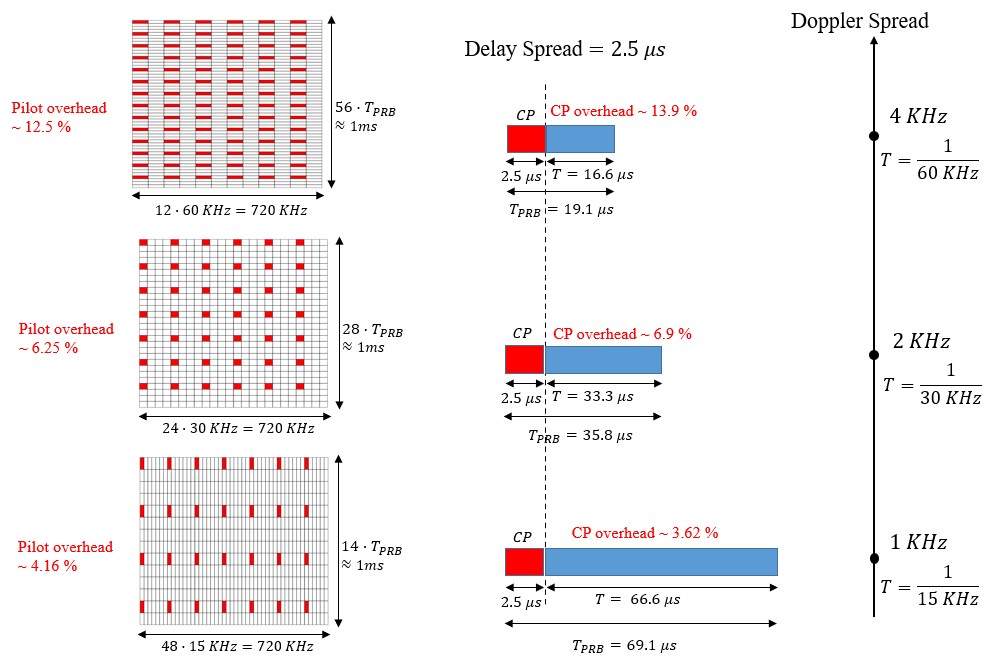}
    	\caption{}
    	\label{fig22}
    \end{subfigure}
    \begin{subfigure}[b]{0.5\textwidth}
	\includegraphics[width=1.0\textwidth]{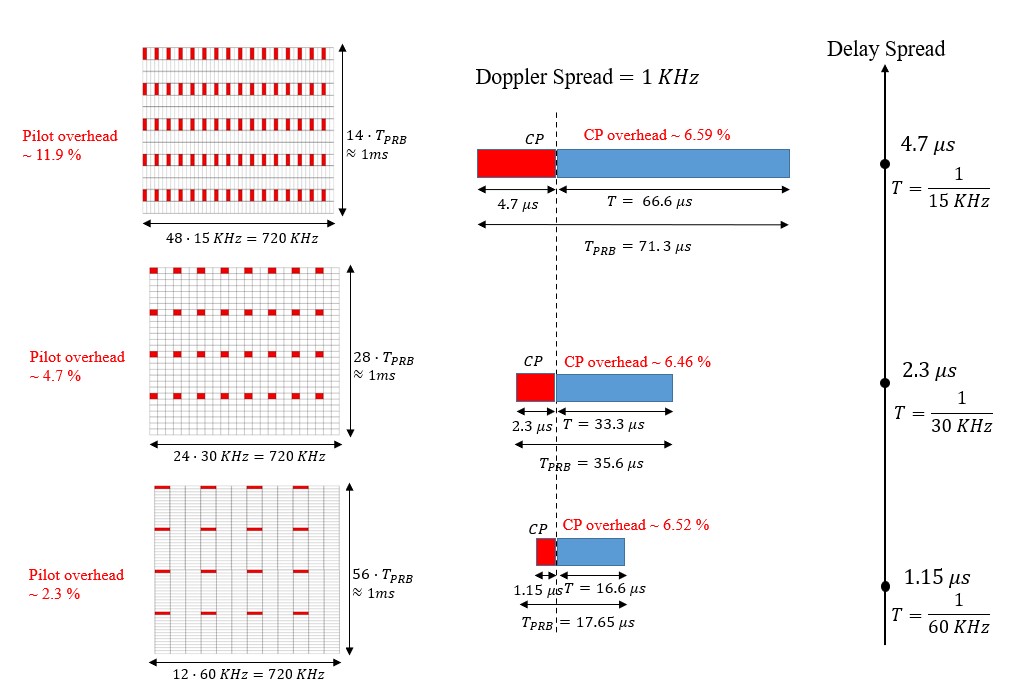}
	\caption{{}}
	\label{fig44}
	\end{subfigure}
	\vspace{-2mm}
	\caption{(a) Effect of channel Doppler spread on overhead in CP-OFDM. (b) Effect of channel delay spread on overhead in CP-OFDM.} 
 \label{fig94}
\end{figure}

\subsection{Cost of preventing ICI}
 Sub-carrier spacing in CP-OFDM increases with Doppler spread to avoid ICI. Hence, the symbol duration decreases but the size of the CP remains the same since it depends only on the channel delay spread.\footnote{\footnotesize{{In 3GPP 5G NR, numerology formats are specified in such a way that the CP overhead is the same. In other words, with every doubling in the sub-carrier spacing the CP duration is halved. The reason argued is that higher sub-carrier spacing would be used for higher carrier frequencies since the channel Doppler spread is higher for the same relative speed between transmitter and receiver and the required CP is smaller since the channel delay spread is smaller due to higher path loss.
 This standard argument for shrinking CP size with increasing sub-carrier spacing does not apply in our discussion here since we consider higher mobility without any increase in the carrier frequency.}}}  Fig.~\ref{fig22} illustrates how this CP overhead increases with Doppler spread. In addition, the channel coherence time decreases as the Doppler Spread increases, so that pilots need to be transmitted more frequently in time. This pilot overhead adds to the CP overhead. Fig.~\ref{fig22} illustrates how the sum of pilot and CP overhead increases from 7.68\% to 28.4\% as the Doppler spread increases from $1$ KHz to $4$ KHz.

\begin{figure}
     \centering
        \vspace{-5mm}
        \includegraphics[width=9.32cm,height=7.0cm]{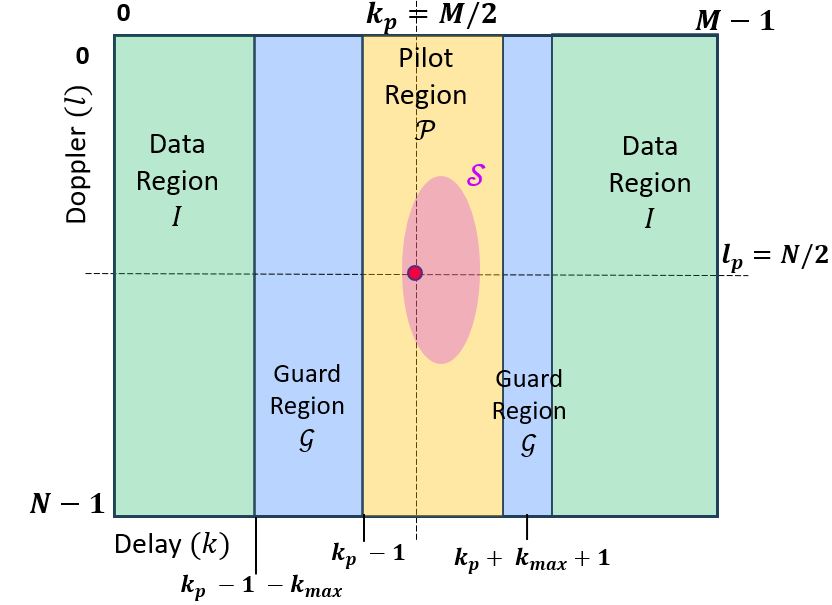}
        \vspace{-2mm}
        \caption{A typical Zak-OTFS frame with pilot, guard and data regions.}
        \label{figframe}
    \end{figure}

Fig.~\ref{figframe} illustrates a Zak-OTFS frame with a point pilot shown as a red dot surrounded by pilot and guard regions. The pilot region is used to sense the I/O relation from the channel response to the point pilot, and the pink ellipse represents the delay and Doppler spread of the pilot response. Data transmission interferes with sensing and the guard regions reduce this interference. The pilot and guard regions represent the overhead in Zak-OTFS with a point pilot. Here we note that this overhead is not present when we consider Zak-OTFS with a spread pilot (see \cite{spreadpilotpaper} for details).

The Doppler period $\nu_p$ is chosen to be greater than the Doppler spread of the pink ellipse, and the delay period $\tau_p$ is chosen to be greater than the delay spread of the pink ellipse. This is the crystallization condition which enables joint equalization of all carriers by eliminating aliasing in the DD domain (see [2, 3] for details). In CP-OFDM the sub-carrier spacing $\Delta f$ is required to be the same for all users. Hence the user with the most significant Doppler spread determines the overhead for all users. \emph{Zak-OTFS is more flexible than CP-OFDM} in that individual users can have different delay and Doppler periods. It is also possible to provision a single delay and Doppler period for all users, and to vary the pilot structure to accommodate different delay and Doppler spreads (see \cite{zakpilots} for more details).

We now consider a high Doppler spread channel where the ratio of the channel Doppler spread to the Doppler spread $1/T$ of an individual Zak-OTFS carrier is greater than the ratio of the channel delay spread to the delay spread $1/B$ of an individual Zak-OTFS carrier. It is natural to choose the pilot and guard regions to be strips along the Doppler axis as shown in Fig.~\ref{figframe}. The length of the strip is the Doppler period $\nu_p$ and this is sufficient for all channels with Doppler spread less than $\nu_p$. The width of the strip depends only on the channel delay spread, it is governed by the geometry of dominant reflectors, it is independent of channel Doppler spread. Unlike CP-OFDM, the overhead of Zak-OTFS does not increase with channel Doppler spread. For example, when the channel delay spread is $2.5 \mu s$ and the channel Doppler spread is $4$ KHz, we can choose the Doppler period $\nu_p= 5$ KHz and the delay period $\tau_p = 1/\nu_p = 200 \mu s$. The guard and pilot regions result in an overhead of $2 \times 2.5/200 = 2.5\%$ which is much smaller than the overhead of $28.4\%$ for CP-OFDM (see Fig.~\ref{fig22}). 

Finally, we consider high delay spread channels where the ratio of the channel Doppler spread to the Doppler spread $1/T$ of an individual Zak-OTFS carrier is smaller than the ratio of the channel delay spread to the delay spread $1/B$ of an individual Zak-OTFS carrier. For a fixed channel Doppler spread, CP size in OFDM would increase with channel delay spread. We can avoid introducing overhead by decreasing sub-carrier spacing, thereby increasing the symbol duration to the point where the CP overhead is unchanged. However, the coherence bandwidth decreases as the channel delay spread increases, making it necessary to introduce additional pilots in the frequency domain. Fig.~\ref{fig44} illustrates the increase in pilot overhead. As the channel delay spread increases from $1.15 \mu s$ to $4.7 \mu s$, the overhead of CP-OFDM increases from $8.8\%$ to $18.5\%$. With Zak-OTFS, we use pilot and guard regions that are strips along the delay axis, and choosing $\tau_p  = 6.25 \mu s$, $\nu_p = 160$ KHz, results in overhead of $2 \times 1/160 = 1.25\%$.
   \begin{figure}
     \centering
        \vspace{-1mm}\includegraphics[width=14.32cm,height=9.7cm]{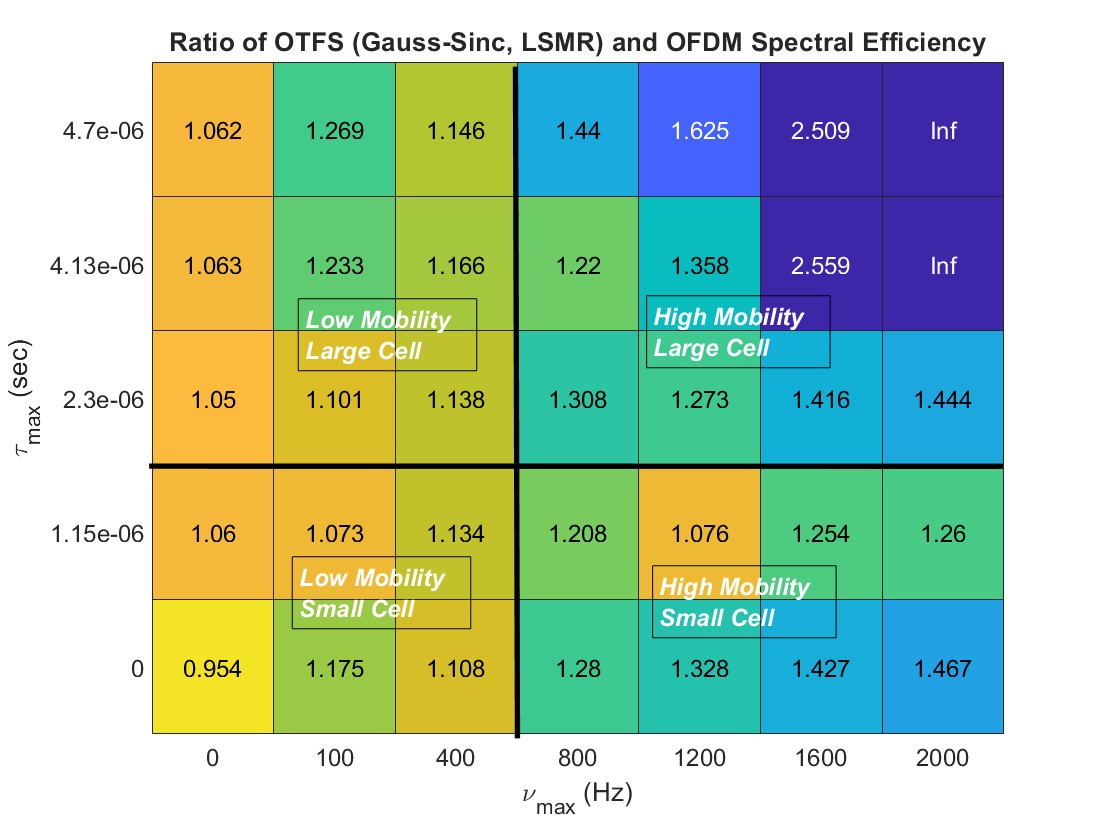}
        \vspace{-2mm}
        \caption{Ratio of effective SE achieved by Zak-OTFS  to that achieved by CP-OFDM.}
        \label{fig_otfsvsofdm}
        \vspace{-3mm}
    \end{figure}
\section{Performance evaluation}
We have argued that 6G propagation environments are changing the balance between time-frequency methods, such as CP-OFDM, and delay-Doppler methods such as Zak-OTFS. In CP-OFDM, once the I/O relation is known, equalization is relatively simple. {However, acquisition of the complete I/O relation in CP-OFDM is nearly impossible}. In Zak-OTFS, equalization is more involved, but complete acquisition of the I/O relation is simple and not dependent on any mathematical model for the underlying physical channel. In this Section we report numerical simulations for a Veh-A channel from 3GPP standards that is representative of 6G propagation environments. 

We consider a single radio resource with finite time duration and bandwidth. We measure the effect of overheads by calculating effective spectral efficiency (SE), which is the ratio of the total number of information bits transmitted reliably divided by the time-bandwidth product. For both CP-OFDM and Zak-OTFS, the time duration is $1$ ms. The resource bandwidth is $720$ KHz for CP-OFDM and $672$ KHz for Zak-OTFS.
We follow 3GPP standards in defining reliable transmission as block error rate (BLER) less than 0.1, though we would note that a $0.1$ BLER introduces significant retransmission overhead. For CP-OFDM, we optimize effective SE over all possible transmission modes in the 3GPP 5G NR standard \cite{3gppmcs1, 3gppmcs2}. In Zak-OTFS, we optimize the effective SE over a small set of Doppler periods $\nu_p$. We calculate the ratio of effective SE for Zak-OTFS to effective SE for CP-OFDM for a given total transmit power to noise ratio, over a discrete set of delay spreads $\tau_{max}$ and Doppler spreads $\nu_{max}$ representing the different types of cellular systems. The total power to noise ratio is $12$ dB, we consider $\tau_{max} = 0, 1.17, 2.34. 4.16, 4.7 \mu s$, and $\nu_{max} = 0, 100, 400, 800, 1200, 1600$, and $2000$ Hz. The numerical simulations result in the heatmap shown in Fig.~\ref{fig_otfsvsofdm}, which is divided into $4$ quadrants representing $4$ types of cellular system - low mobility/small cell, high mobility/small cell, low mobility/large cell, and high mobility/large cell. 

For both CP-OFDM and Zak-OTFS, we optimize over modulation and coding schemes (MCS) specified in the 3GPP 5G NR standard \cite{3gppmcs1, 3gppmcs2}. For CP-OFDM we optimize SE over all combinations of MCS, Type-A DMRS pilot allocation, DMRS power boost of $-6, -4, -2, 0, 2, 4, 6$ dB relative to the power on each data sub-carrier, and sub-carrier spacing $\Delta f = 15, 30$, and $60$ KHz. We select the combination that maximizes effective SE (as defined above) with a total transmit power to noise ratio of $12$ dB. {For OFDM we consider per sub-carrier MMSE equalization.}

\begin{figure}
     \centering
        \vspace{-1mm}\includegraphics[width=17.32cm,height=6.0cm]{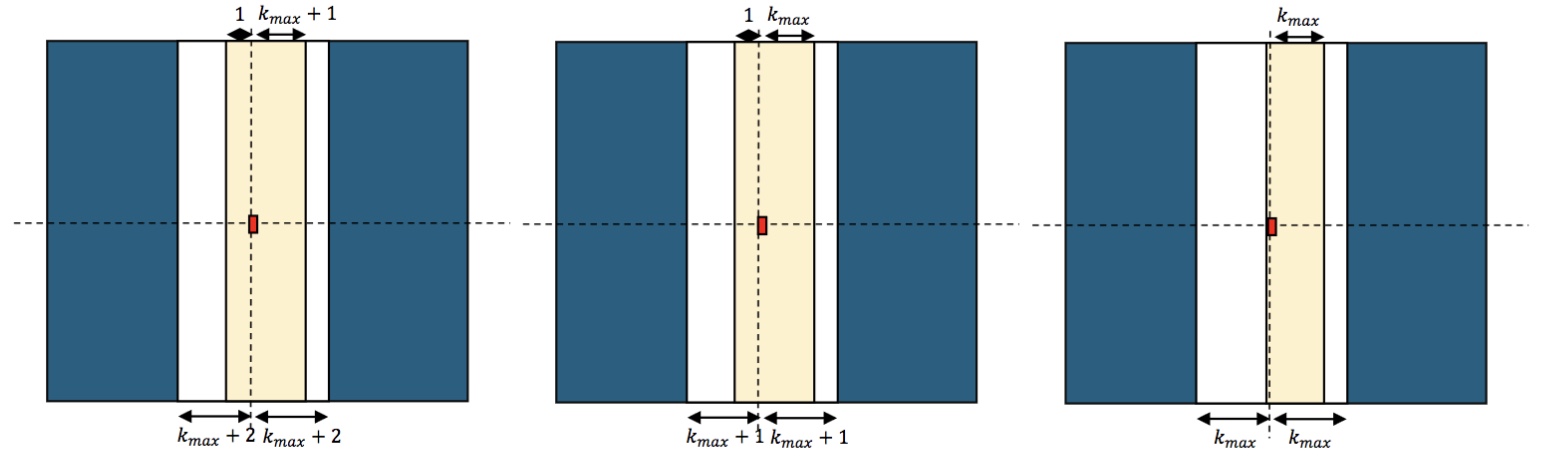}
        \vspace{-2mm}
        \caption{{Three different Zak-OTFS frame structures for pilot-data allocation. Each frame is $M$ taps wide along the horizontal delay axis and is $N$ taps wide along the vertical Doppler axis. In each frame structure, the red dot in the middle depicts the pilot pulsone used for acquiring the effective discrete DD domain channel filter.
        The received pulsones in the yellow colored ``pilot region" are used for this estimation. The width of this pilot region should be at least $k_{max} = \lceil B \tau_{max} \rceil$ taps in order that a significant fraction of the received pilot energy is used for channel estimation. The pulsones in the gray colored ``data region" carry information and those in the white colored ``guard region" are used to avoid interference between the data and the pilot pulsones. The three different frame structures differ in the width of the pilot region. A wider pilot region results in more accurate acquisition of the effective DD domain channel filter and reduced interference between data and pilots, but as the cost of higher overhead since the pulsones in the guard and pilot region do not carry information.}}
        \label{fig_pilotalloc}
        \vspace{-3mm}
    \end{figure}

{For Zak-OTFS, we optimize over all combinations of MCS, Doppler period $\nu_p=1,2,4,6, 8, 12, 14, 24$ KHz, pilot to data power ratio (PDR) $-15, -10, -5, 0, 5, 10, 15$ dB, and different pilot-data allocations (shown in Fig.~\ref{fig_pilotalloc}. Again, we select the combination that maximizes effective SE with a total transmit power to noise ratio of $12$ dB. We use Gauss-Sinc pulse shaping filters (see \cite{Gaussinc}) and detect Zak-OTFS information symbols from the received DD domain samples using the LSMR joint equalizer \cite{lsmreq}. }
In Fig.~\ref{fig_otfsvsofdm}, we observe that for high mobility/large cell systems such as non-terrestrial networks (NTN) \cite{NTN} and aircraft-to-ground communications \cite{airtoground}, the effective SE of Zak-OTFS is more than double that of CP-OFDM. In high mobility/small cell systems such as high speed trains and cars at highway speeds Zak-OTFS improves effective SE by more than $20\%$ in most cases. Zak-OTFS also improves effective SE in low mobility/large cell systems typical of rural connectivity in large parts of India, Australia, Africa and South America (the LMLC scenario in 3GPP 5G NR \cite{LMLC}).

{In Fig.~\ref{fig_otfsvsofdm}, we also observe
that for low mobility and small cell scenarios (i.e., low delay and Doppler spread) the SE achieved by both CP-OFDM and OTFS is similar. In this scenario there is insignificant ICI in CP-OFDM and therefore low-complexity per sub-carrier equalization is nearly optimal.}

\subsection{Choice of operating point}
When 3GPP defines reliable transmission by specifying a block error rate of $0.1$, it is selecting an operating point where noise dominates ICI. This limits the benefit to acquiring the I/O relation. The cost of this choice is significant overhead in retransmission, and this price is paid above the physical layer. 

Why not set the BLER to $0.01$? Is it because CP-OFDM is not able to achieve positive SE at this BLER because of ICI at high Doppler spreads? Fig.~\ref{fig_otfsvsofdm} illustrates that this is the case for $\nu_{max} = 2000$ Hz and $\tau_{max} = 4.7 \mu s$ where CP-OFDM is not able to achieve BLER $0.1$ for any MCS combination. {A BLER operating point lower than $0.1$ is desirable as it can result in a higher effective SE due to reduced re-transmissions.}

We have seen that CP overhead increases with sub-carrier spacing, and the 3GPP 5G NR standard limits this overhead by reducing CP by half when the sub-carrier spacing doubles. This feature of the standard is not motivated by high mobility or higher RF carrier frequencies at FR2, where channel Doppler spread could increase without any effect on channel delay spread. 

\section{Conclusions}
In the full paper we will provide a system level view of the architectural choice between preventing ICI and embracing ICI. Our objective is to provide theoretical insight into the impact on system performance of choices made at the physical layer. We will explore the potential benefits that might accrue from exploring different operating points in cellular standards.


\begin{thebibliography}{1}
\bibitem{IMT2030}
``Framework and Overall Objectives of the Future Development of IMT for 2030 and Beyond," Recommendation M.2160-0, International Telecommunication Union (Radiocommunication), Nov. 2023. 
\bibitem{zakotfs1}
S. K. Mohammed, R. Hadani, A. Chockalingam, and R. Calderbank, ``OTFS – A mathematical foundation for communication and radar sensing in the delay-Doppler domain,'' {\it IEEE BITS the Information Theory Magazine}, vol. 2, no. 2, pp. 36-55, 1 Nov. 2022.

\bibitem{zakotfs2}
S. K. Mohammed, R. Hadani, A. Chockalingam and R. Calderbank, ``OTFS—Predictability in the Delay-Doppler Domain and Its Value to Communication and Radar Sensing," in IEEE BITS the Information Theory Magazine, vol. 3, no. 2, pp. 7-31, June 2023.

\bibitem{otfsbook}
{S. K. Mohammed, R. Hadani and A. Chockalingam, {\it OTFS Modulation: Theory and Applications}, IEEE Press and Wiley, Nov. 2024.}

\bibitem{Bestreadings2022}
``Best Readings in Orthogonal Time Frequency Space (OTFS) and Delay Doppler Signal Processing,” June 2022. https://www.comsoc.org/publications/best-readings/orthogonal-time-frequency-space-otfs-and-delay-doppler-signal-processing

\bibitem{OFDM1971}
S. Weinstein and P. Ebert, ``Data Transmission by Frequency-Division Multiplexing Using the Discrete Fourier Transform," in IEEE Transactions on Communication Technology, vol. 19, no. 5, pp. 628-634, October 1971.


\bibitem{Wang2006}
T. Wang, J. G. Proakis, E. Masry and J. R. Zeidler, ``Performance Degradation of OFDM Systems Due to Doppler Spreading,” IEEE Trans. on Wireless Commun., vol. 5, no. 6, June 2006.


\bibitem{zakpilots}
J. Jayachandran, I. A. Khan, S. K. Mohammed, R. Hadani, A. Chockalingam, R. Calderbank, ``Zak-OTFS with Interleaved Pilots to Extend the Region of Predictable Operation," arXiv preprint arXiv:2408.09379v2, Dec 2024.


\bibitem{EVAITU} 
ITU-R M.1225, ``Guidelines for evaluation of radio transmission technologies for IMT-2000,'' {\it International Telecommunication Union Radio communication}, 1997.

\bibitem{3gppmcs1}
3GPP TS 38.211, ``NR; Physical Channels and Modulation," Release 15, 2018.

\bibitem{3gppmcs2}
3GPP TS 38.212, ``NR; Multiplexing and Channel Coding," Release 15, 2018.

\bibitem{NTN}
A. Guidotti et al., ``Role and Evolution of Non-Terrestrial Networks Toward 6G Systems," \emph{IEEE Access}, vol. 12, pp. 55945-55963, 2024.

\bibitem{Gaussinc}
A.~Das, F.~Jesbin and A.~Chockalingam, ``A Gaussian-Sinc Pulse Shaping Filter for Zak-OTFS," under review (available online, arXiv:2502.03904 [cs.IT]), Feb. 2025.

\bibitem{lsmreq}
H. Qu, G. Liu, L. Zhang, S. Wen and M. A. Imran, ``Low-Complexity Symbol Detection and Interference Cancellation for OTFS System," \emph{IEEE Transactions on Communications}, vol. 69, no. 3, pp. 1524-1537, March 2021.

\bibitem{airtoground}
W. Khawaja, I. Guvenc, D. W. Matolak, U. -C. Fiebig and N. Schneckenburger, ``A Survey of Air-to-Ground Propagation Channel Modeling for Unmanned Aerial Vehicles," \emph{IEEE Communications Surveys \& Tutorials}, vol. 21, no. 3, pp. 2361-2391, thirdquarter 2019.

\bibitem{LMLC}
M. P. Reddy, G. K. Rao, D. H. Kumar, K. Subhash, S. Amuru and K. Kuchi, ``Uplink coverage enhancements
for extremely large‑cell sites," \emph{Eurasip Journal on Wireless Communications and Networking}, 2022.

\bibitem{pulseshaping}
J. Jayachandran, R. K. Jaiswal, S. K. Mohammed, R. Hadani, A. Chockalingam and Robert Calderbank, ``Zak-OTFS: Pulse Shaping and the Tradeoff between Time/Bandwidth Expansion and Predictability," arXiv preprint arXiv:2405.02718, 2024.

\bibitem{spreadpilotpaper}
M. Ubadah, S. K. Mohammed, R. Hadani, S. Kons, A. Chockalingam and R. Calderbank, ``Zak-OTFS to Integrate Sensing the I/O Relation and Data Communication," submitted to IEEE Transactions on Information Theory, January 2025.
\end{thebibliography}
\end{document}